\pdfoutput=1
\documentclass[11pt, a4paper]{scrartcl}
\usepackage[l2tabu, orthodox]{nag}
\usepackage[utf8]{inputenc}
\usepackage[pdfusetitle]{hyperref}

\usepackage{amsmath, amsthm, amssymb, amsfonts}

\usepackage{algorithmic}
\usepackage{algorithm}
\usepackage{subcaption}
\usepackage{mathptmx}
\usepackage[scaled=.90]{helvet}

\usepackage{graphicx}
\usepackage{authblk}

\DeclareMathOperator{\Adj}{Adj}

\usepackage{float}

\usepackage[semicolon, round]{natbib}

\usepackage{doi}
\begin{document}

\author[1,*]{Gheorghe-Teodor~Bercea}
\author[2,3,4]{Andrew~T.~T.~McRae}
\author[3]{David~A.~Ham}
\author[1,3]{Lawrence~Mitchell}
\author[1,5]{Florian~Rathgeber}
\author[1]{Luigi~Nardi}
\author[1]{Fabio~Luporini}
\author[1]{Paul~H.~J.~Kelly}

\affil[1]{Department of Computing, Imperial College London, London, SW7 2AZ, UK}
\affil[2]{The Grantham Institute, Imperial College London, London, SW7 2AZ, UK}
\affil[3]{Department of Mathematics, Imperial College London, London, SW7 2AZ, UK}
\affil[4]{Department of Mathematical Sciences, University of Bath, Bath, BA2 7AY, UK}
\affil[5]{European Centre for Medium-Range Weather Forecasts, Reading, RG2 9AX, UK}
\affil[*]{Correspondance to: \texttt{gb308@doc.ic.ac.uk}}

\title{A structure-exploiting numbering algorithm for finite elements on
  extruded meshes, and its performance evaluation in Firedrake}
\date{}

\maketitle

\begin{abstract}
  We present a generic algorithm for numbering and then efficiently
  iterating over the data values attached to an extruded mesh. An
  extruded mesh is formed by replicating an existing mesh, assumed to
  be unstructured, to form layers of prismatic cells. Applications of
  extruded meshes include, but are not limited to, the representation
  of 3D high aspect ratio domains employed by geophysical finite
  element simulations. These meshes are structured in the extruded
  direction. The algorithm presented here exploits this structure to
  avoid the performance penalty traditionally associated with
  unstructured meshes. We evaluate the implementation of this
  algorithm in the Firedrake finite element system on a range of low
  compute intensity operations which constitute worst cases for data
  layout performance exploration. The experiments show that having
  structure along the extruded direction enables the cost of the
  indirect data accesses to be amortized after 10-20 layers as long as
  the underlying mesh is well-ordered. We characterise the resulting
  spatial and temporal reuse in a representative set of both
  continuous-Galerkin and discontinuous-Galerkin discretisations. On
  meshes with realistic numbers of layers the performance achieved is
  between 70\% and 90\% of a theoretical hardware-specific limit.
\end{abstract}

\section{Introduction}
\label{sec:intro}

In the field of numerical simulation of fluids and structures, there
is traditionally considered to be a tension between the computational
efficiency and ease of implementation of structured grid models, and
the flexible geometry and resolution offered by unstructured meshes.

In particular, one of the grand challenges in simulation science is
modelling the ocean and atmosphere for the purposes of predicting the
weather or understanding the Earth's climate system. The current
generation of large-scale operational atmosphere and ocean models
almost all employ structured meshes \citep{Slingo:2009}. However,
requirements for geometric flexibility as well as the need to overcome
scalability issues created by the poles of structured meshes has led
in recent years to a number of national projects to create
unstructured mesh models \citep{Ford:2013,Zaengl:2015,Skamarock:2012}.

The ocean and atmosphere are thin shells on the Earth's surface, with
typical domain aspect ratios in the thousands (oceans are a few
kilometres deep but thousands of kilometres across). Additionally the
direction of gravity and the stratification of the ocean and
atmosphere create important scale separations between the vertical and
horizontal directions. The consequence of this is that even
unstructured mesh models of the ocean and atmosphere are in fact only
unstructured in the horizontal direction, while the mesh is composed
of aligned layers in the vertical direction. In other words, the
meshes employed in the new generation of models are the result of
extruding an unstructured two-dimensional mesh to form a layered mesh
of prismatic elements.

This layered structure was exploited in \citet{Macdonald:2011}
to create a numbering for a finite volume atmospheric model
such that iteration from one cell to the next within a vertical column
required only direct addressing.  They show that when only paying the
price of indirect addressing on the base mesh there is less than 5\% performance
difference between two implementations of an atmospheric model which treat
the same icosahedral mesh first as fully structured and then as partially
structured (extruded). One of the caveats of that comparison is that the
underlying mesh is fully structured in both cases which presents an
advantage to the indirect addressing scheme which is not present for more
general unstructured meshes.

Exploiting the anisotropic nature of domains has seen various software
developments in various fields.  For example \texttt{p6est}
(\citet{Isaac:2015} and \citet[\S 2.3]{Isaac:2015a}), a package for 2+1
dimensional adaptive mesh refinement, was developed to maintain
columnwise numbering for numerical reasons in ice sheet modelling, but
does not support general unstructured base meshes.  The DUNE-PrismGrid
module~\citep{Gersbacher:2012} provides extruded meshes for any base
DUNE grid, but does not describe a degree of freedom numbering or
provide detailed performance characteristics of the iteration on
extruded meshes.  The Model for Prediction Across Scales (MPAS) uses a
column innermost numbering for their C-grid atmospheric and ocean
model \citep{Sarje:2015}.  Their implementation is limited to the
single discretisation employed by that model.

A key motivation for this work was to provide an efficient mechanism
for the implementation of the layered finite element numerics which
have been adopted by the UK Met Office's Gung Ho programme to develop
a new atmospheric dynamical core. The algorithms here have been
adopted by the Met Office for this purpose \citep{Ford:2013}.  While
geophysical applications motivate this work, the algorithms and their
implementation in Firedrake \citep{Rathgeber:2016} are more general and
could be applied to any high aspect ratio domain.

\subsection{Contributions}
\label{ssec:contributions}
\begin{itemize}
\item We generalize the numbering algorithm in \citet{Macdonald:2011} to the
  full range of finite element discretisations.
\item We demonstrate the effectiveness of the algorithm with respect
  to absolute hardware performance limits.
\end{itemize}

\section{Unstructured Meshes}
\label{sec:unstructured-meshes}

In this section we briefly restate the data model for unstructured
meshes introduced in \citet{Logg:2009,Knepley:2009}. In
\autoref{ssec:mesh-representation} we rigorously define a \emph{mesh},
explain mesh topology, geometry and numbering. In
\autoref{ssec:mesh-data} we explain how data may be associated with
meshes.

\subsection{Terminology}
\label{ssec:mesh-terminology}
When describing a mesh, we need some way of specifying the neighbours
of a given entity.  This is always possible using \emph{indirect
  addressing} in which the neighbours are explicitly enumerated, and
sometimes possible with \emph{direct addressing} where a closed form
mathematical expression suffices.

In what follows we start with a \emph{base mesh} which we will
\emph{extrude} to form a mesh of higher topological dimension.  Due to
geophysical considerations, we refer to the plane of the base mesh as
the \emph{horizontal} and to the layers as the \emph{vertical}.

We will also employ the definition of a \emph{graph} as a set $V$ and
a set $E$ of edges where each edge represents the relationships
between the elements of the set $V$.

\subsection{Meshes}
\label{ssec:mesh-representation}

A mesh is a decomposition of a simulation domain into non-overlapping
polygonal or polyhedral cells. We consider meshes used in algorithms
for the automatic numerical solution of partial differential
equations. These meshes combine topology and geometry. The topology of
a mesh is composed of mesh entities (such as vertices, edges, cells)
and the adjacency relationships between them (cells to vertices or
edges to cells). The geometry of the mesh is represented by
coordinates which define the position of the mesh entities in space.

Every mesh entity has a topological dimension given by the minimum
number of spatial dimensions required to represent that entity. We
define $D$ to be the minimum number of spatial dimensions needed to
represent a mesh and all its entities. A vertex is representable in
zero-dimensional space, similarly an edge is a one-dimensional entity
and a cell a $D$-dimensional entity. In a two-dimensional mesh of
triangles, for example, the entities are the vertices, edges and
triangle cells with topological dimensions $0$, $1$ and $2$
respectively. The minimum number of geometric dimensions needed to
represent the mesh and all its entities is $D=2$.

A mesh can be represented by several graphs. Each graph consists of a
multi-type set $V$ and a typed adjacency relationship
$\Adj_{d_{1}, d_{2}}$ between $d_{1}$- and $d_{2}$-typed elements in
$V$. The type of an entity in $V$ is simply its dimension. The
adjacency graphs will always map from a set of uniform dimension to a
set of uniform dimension.  Attaching types to elements of $V$ enables
graphs to capture the relationships between different mesh entities,
for example cells and vertices, edges and vertices.

We write $V_{d}$ to mean the set of mesh entities of topological
dimension $d$ where $0 \leq d \leq D$:
\begin{equation}
V_{d}=\{ (d, i)\ |\ 0 \leq i \leq N_{d} - 1\},
\end{equation}
where $N_{d}$ is the number of entities of dimension $d$. The set $V$
is then simply the union of the $V_d$s:
\begin{equation}
V =\bigcup_{0 \leq d \leq D} V_{d}.
\end{equation}

Every mesh entity has a number of adjacent entities. The mesh-element
connectivity relationships are used to specify the way mesh entities
are connected. For a given mesh of topological dimension $D$ there are
$(D+1)^2$ different types of adjacency relationships. To define the
mesh, only a minimal subset of relationships from which all the others
can be derived is required.  For example, as shown in
\citet{Logg:2009}, the complete set of adjacency relationships may be
derived from the cell-vertex adjacency.

We write
\begin{equation}
 \Adj_{d_{1}, d_{2}}(v) = (v_{1}, v_{2}, \dots, v_{k}),
\end{equation}
to specify the entities $v_1, v_2, \dots, v_k \in V_{d_2}$ adjacent to
$v \in V_{d_1}$.

In a mesh with a very regular topology, there may be a closed form
mathematical expression for the adjacency relationship
$\Adj_{d_{1}, d_{2}}(v)$. Such meshes are termed
\emph{structured}. However since we are also interested in supporting
more general \emph{unstructured meshes}, we must store the lists of
adjacent entities explicitly.

\subsection{Attaching data to meshes}
\label{ssec:mesh-data}

Every mesh entity has a number of values associated with it. These
values are also known as \emph{degrees of freedom} and they are the
discrete representation of the continuous data fields of the domain.
As the degrees of freedom are uniquely associated with mesh entities,
the mesh topology can be used to access the degrees of freedom local
to any entity using the connectivity relationships.

A \emph{finite element discretisation} associates a number of degrees
of freedom with each entity of the mesh. A \emph{function space} uses
the discretisation to define a numbering for all the degrees of
freedom. Multiple different function spaces may be defined on a mesh
and each function space may have several data fields associated with
it. In the case of a triangular mesh for example, a piecewise linear
function space will associate a degree of freedom with every vertex of
the mesh while a cubic function space will associate one degree of
freedom with every vertex, two degrees of freedom with every edge and
one degree of freedom with every cell. In the former case there will
be three degrees of freedom adjacent to a cell, and a total of ten in
the latter case.

The data associated with the mesh also needs to be numbered.  The
choice of numbering can have a significant effect on the computational
efficiency of calculations over the mesh
\citep{Gunther:2006,Lange:2016,Yoon:2005}.

\subsection{Kernels and stencils}
\label{ssec:kernels-stencils}

The most common operation performed on meshes is the local application
of a function or \emph{kernel} while traversing, or \emph{iterating}
over a homogeneous subset of mesh entities.  The kernel is executed
once for each such mesh entity and acts on the degrees of freedom in a
\emph{stencil} composed of the mesh entities adjacent to the the
iterated entity.  For example, a finite element operator evaluating an
integral over the domain would iterate over the mesh cells and access
data through a stencil comprising the degrees of freedom on that cell
and its adjacent facets, edges, and vertices.  For a more in-depth
discussion on the construction of stencils on unstructured meshes, the
reader is referred to \cite{Logg:2009} and \cite{Knepley:2009}.  In
theory, this requires cell-to-facets, cell-to-edges, and
cell-to-vertices adjacency relationships (cell-to-cell is
implicit). In practice the three different relationships may be
composed into a single adjacency relationship which references the
data associated with all the different adjacent entity types.

In the unstructured case, we store an explicit list (also known as a
\textit{map}) $L(e)$ for each type of stencil operation which given a
topological entity $e$ returns the set of degrees of freedom in the
stencil at that entity.

\section{Extruded Meshes}
\label{sec:extruded-meshes}

In \autoref{ssec:extruded-mesh-definition} we introduce extruded
meshes and in \autoref{ssec:extruded-mesh-definition} we show how the
entities and the data are to be numbered. In
\autoref{ssec:ext-mesh-iteration} we present the extruded mesh
iteration algorithm and the offset computation for the direct
addressing scheme along the vertical direction.

\subsection{Definition of an Extruded Mesh}
\label{ssec:extruded-mesh-definition}

An extruded mesh consists of a base mesh which is replicated a fixed
number of times in a layered structure\footnote{For ease of
  exposition, we discuss the case where each mesh column contains the
  same number of layers, however this is not a limitation of the
  method and algorithms presented here}. A mesh of topological
dimension $D$ becomes an extruded mesh of topological dimension $D+1$.

The mesh definition can be extended to include extruded meshes. Let
mesh $M = (V, \Adj)$ be a non-extruded mesh where $\Adj$ stands for
all the valid adjacency relationships of $M$. An extruded mesh which
has $M$ as the base mesh can be defined as a triple
$(V^\textrm{extr}, \Adj^\textrm{extr}, \lambda)$ where
$\Adj^\textrm{extr}$ is the set of valid adjacency relationships and
$\lambda \in \mathbb{N}^+$ is the number of intervals over which the
mesh is extruded.  This implies that there are $\lambda+1$ vertices in
the extruded direction.  Before we can define $V^\textrm{extr}$ and
$\Adj^\textrm{extr}$ several concepts have to be introduced.

\subsubsection{Tensor product cells}
\label{sssec:tensor-product-cells}
The effect of the extrusion process on the base mesh can always be
captured by associating a line segment with the vertical direction. We
write $D_{b}$ for the topological dimension of the base mesh while the
topological dimension of the vertical mesh is always equal to $1$.

As a consequence, the cells of the extruded mesh are prisms formed by
taking the tensor product of the base mesh cell with the vertical line
segment. For example, each triangle becomes a triangular prism.  The
construction of tensor product cells and finite element spaces on them
is considered in more detail in \citet{McRae:2016}.

\subsubsection{Extruded Mesh Entities}
\label{sssec:extruded-mesh-entities}

\begin{figure}[htbp]
\centering
\includegraphics[width=.6\linewidth]{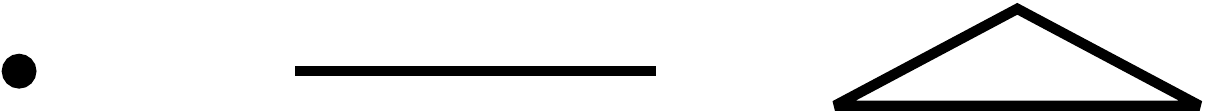}
\caption{Extruded mesh entities belonging to the base mesh to be
  extruded (left to right): vertices, horizontal edges, horizontal
  facets.}
\label{fig:base-mesh-entities}
\end{figure}

\begin{figure}[htbp]
\centering
\includegraphics[width=.6\linewidth]{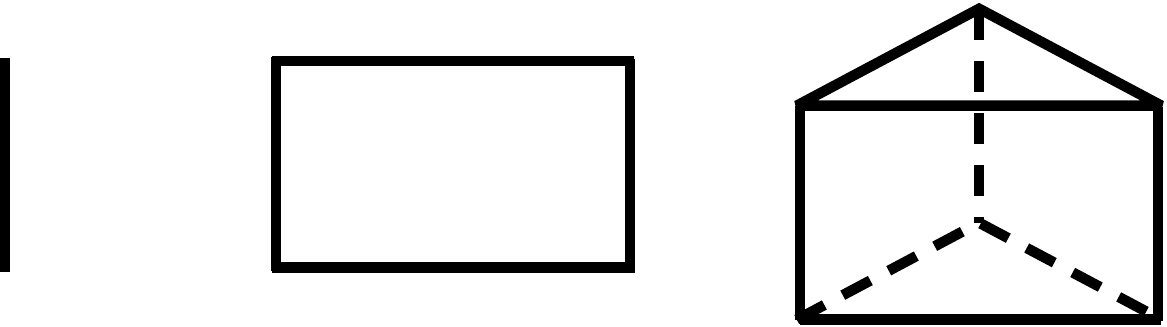}
\caption{Mesh entities used in the extrusion process to connect
  entities in \autoref{fig:base-mesh-entities} (left to right):
  vertical edges, vertical facets, 3D cells.}
\label{fig:ext-mesh-entities}
\end{figure}

The extrusion process introduces new types of mesh entities reflecting
the connectivity between layers. The pairs of corresponding entities
of dimension $d$ in adjacent layers are connected using entities of
dimension $d+1$. In a triangular mesh for example
(\autoref{fig:base-mesh-entities}), the corresponding vertices are connected
using vertical edges, edges contained in each layer are connected by
quadrilateral facets and the 2D triangle faces are connected by a 3D
triangular prism (\autoref{fig:ext-mesh-entities}).

The topological dimension on its own is no longer enough to
distinguish between the different types of entities and their
orientation. Instead entities are characterised by a pair composed of
the horizontal and vertical dimensions. In the case of a 2D triangular
base mesh the set of dimensions is $\{0, 1, 2\}$. The line segment of
the vertical can be described by the set of dimensions $\{0, 1\}$. The
Cartesian product of the two sets yields a set of pairs
(\autoref{eq:entity-pairs}) which can be used to uniquely identify
mesh entities.
\begin{equation}
\label{eq:entity-pairs}
\{ (0, 0), (0, 1), (1, 0), (1, 1), (2, 0), (2, 1) \}
\end{equation}
We refer to the components of each pair as the \textit{horizontal} and
\textit{vertical} dimension of the entity respectively.
\autoref{tab:ext-mesh-dimensions} shows the mapping between the mesh
entity types and their descriptor.
\begin{table}[htbp]
  \centering
  \begin{tabular}{l|l}
    \emph{Mesh entity} & \emph{Dimensions} \\
    \hline
    Vertex             & $(0, 0)$          \\
    Vertical Edge      & $(0, 1)$          \\
    Horizontal Edge    & $(1, 0)$          \\
    Vertical Facet     & $(D_{b}-1, 1)$    \\
    Horizontal Facet   & $(D_{b}, 0)$      \\
    Cell               & $(D_{b}, 1)$      \\
  \end{tabular}
  \caption{Topological dimensions of extruded mesh entities. $D_{b}$
    denotes the topological dimension of the base
    mesh.\label{tab:ext-mesh-dimensions}}
\end{table}

\subsubsection{Extruded Mesh Entity Numbering}
\label{sssec:entitynumbering}
We write $V_{d_{1}, d_{2}}$ to denote the set of topological entities
which are the tensor product of entities of dimensions $d_{1}$ in the
horizontal and $d_{2}$ in the vertical ($0 \leq d_{1} \leq D_{b}$ and
$0 \leq d_{2} \leq 1$):
\begin{equation}
V_{d_{1}, d_{2}}=\{ ((d_{1}, d_{2}), (i, l))\ |\  0 \leq i \leq N_{d_{1}} - 1,\ 0\leq l \leq \lambda - d_{2}\},
\end{equation}
where $N_{d_{1}}$ is the number of entities of dimension $d_{1}$ in
the base mesh and $\lambda$ is the number of edges in the extruded
direction. The subtraction of $d_{2}$ from $\lambda$ accounts for the
fencepost error caused by the fact that there is always one fewer edge
than vertex in the vertical direction.

\begin{figure}[htbp]
  \centering
  \includegraphics[width=14cm]{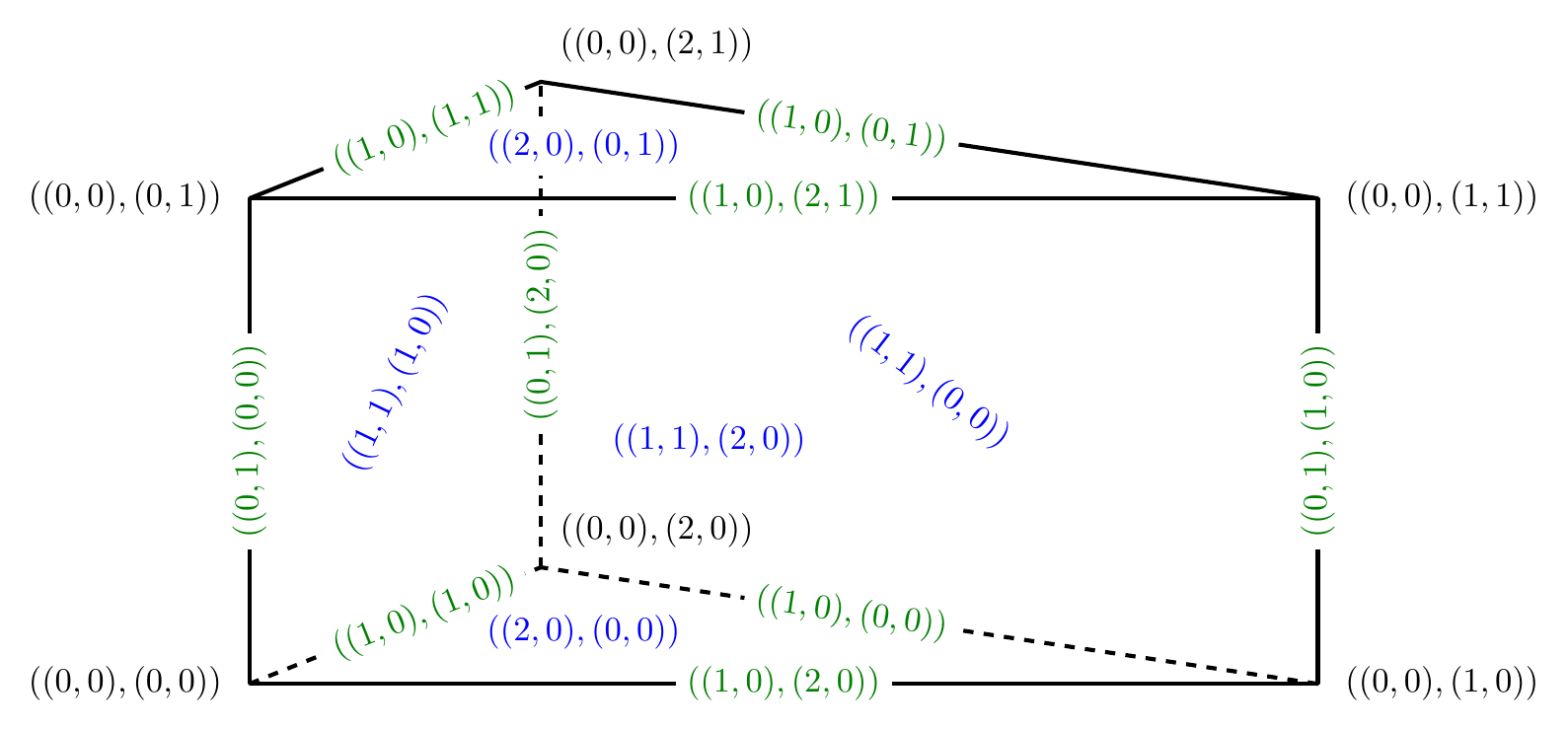}
  \caption{Numbering of the topological entities of an extruded cell
    for the case of an extruded triangle.  The cell itself has
    numbering $((2, 1), (0, 0))$ (not shown), the other entities are
    numbered as shown with vertices in black, edges in green, and faces
    in blue.}
  \label{fig:extruded-wedge}
\end{figure}

The complete set of extruded mesh entities is then
\begin{equation}
V^\textrm{extr} =\bigcup\limits_{\substack{ 0 \leq d_{1} \leq D_{b} \\ 0 \leq d_{2} \leq 1}} V_{d_{1}, d_{2}}.
\end{equation}
These entities are the drawn for the case of an extruded triangle in
\autoref{fig:extruded-wedge}.

Similarly we must extend the indexing of the adjacency relationships,
writing:
\begin{equation}
 \Adj^\textrm{extr}_{(d_{1}, d_{2}), (d_{3}, d_{4})}(v) = (v_{1}, v_{2}, \dots, v_{k}),
\end{equation}
where $v \in V_{d_{1}, d_{2}}$ and
$v_{1}, v_{2}, \dots, v_{k} \in V_{d_{3}, d_{4}}$.

\subsection{Attaching data to extruded meshes}
\label{ssec:ext-mesh-data}

Identically to the case of non-extruded meshes, function spaces over
an extruded mesh associate degrees of freedom with the (extended) set
of mesh entities. A constant number of degrees of freedom is
associated with each entity of a given type.

If we can arrange that the degrees of freedom are numbered such that
the vertical entities are ``innermost'', it is possible to use direct
addressing for the vertical part of any mesh iteration, significantly
reducing the computational penalty introduced by using an indirectly
addressed, unstructured base mesh.  Algorithm~\ref{alg:one} implements
this ``vertical innermost'' numbering algorithm. The critical feature
of this algorithm is that degrees of freedom associated with
vertically adjacent entities have adjacent global numbers.
\begin{algorithm}[htbp]
\caption{Computing the global numbering for degrees of freedom on an extruded mesh}
\label{alg:one}
\begin{algorithmic}
\REQUIRE {$V$ : the set of base mesh entities}
\REQUIRE {$\lambda$ : the number of vertical intervals}
\REQUIRE {$\delta((d_1, d_2))$ : the number of DoFs associated with each $(d_1, d_2)$ entity}
\ENSURE{$\mathsf{dofs}_{\mathsf{fs}}$: the degrees of freedom associated with each entity}
\STATE $c \leftarrow 0$
\COMMENT{Loop over base mesh entities}
\FOR{$(d_{1}, i)$ in $V$}
        \STATE \COMMENT{Loop over layers}
        \FOR{$l$ in $\{0, 1, ... , \lambda - 1\}$}
            \STATE \COMMENT{Number the horizontal layer, then the connecting entity above it}
            \FOR{$d_{2}$ in $\{0, 1\}$}
                      \STATE \COMMENT{Assign the next $\delta((d_{1}, d_{2}))$ global DoF numbers to this entity}
				\STATE $\mathsf{dofs}_{\mathsf{fs}}((d_{1}, d_{2}), (i, l)) \leftarrow c, c + 1, ..., c + \delta((d_{1}, d_{2})) - 1$

				\STATE $c \leftarrow c + \delta((d_{1}, d_{2}))$
             \ENDFOR
        \ENDFOR
        \STATE \COMMENT{Number the top horizontal layer of this column}
	  \STATE $\mathsf{dofs}_{\mathsf{fs}}((d_{1}, 0), (i, \lambda)) \leftarrow c, c + 1, ..., c + \delta((d_{1}, 0)) - 1$

	  \STATE $c \leftarrow c + \delta((d_{1}, 0))$
\ENDFOR
\end{algorithmic}
\end{algorithm}

\begin{figure}[htbp]
\centering
\includegraphics[height=.35\textheight]{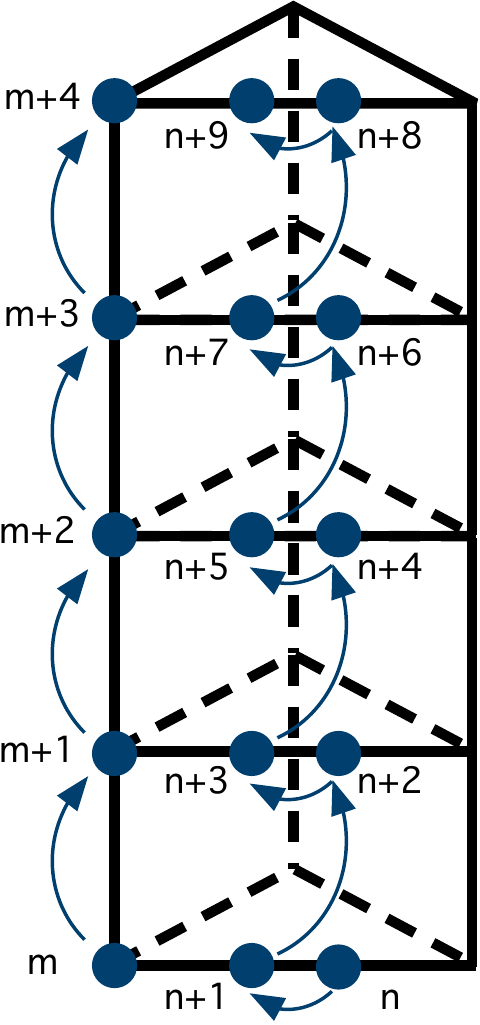}
\caption{Vertical numbering of degrees of freedom (shown in filled
  circles) associated with vertices and horizontal edges. Only one set
  of vertically aligned degrees of freedom of each type is shown. The
  arrows outline the order in which the degrees of freedom are
  numbered.}
\label{fig:vertical-numbering}
\end{figure}

The outcome
of this vertical numbering is shown in
\autoref{fig:vertical-numbering}.  The global numbering algorithm is
orthogonal to any base mesh decomposition strategy used to support
execution on distributed memory parallel systems.  The numbering order
within each entity column is not unique, for example one could
interchange the $l$ and $d_2$ loops in Algorithm~\ref{alg:one}.
However, our choice maximises cache-line usage on a per-element basis.

\subsection{Iterating over extruded meshes}
\label{ssec:ext-mesh-iteration}

Iterating over the mesh and applying a kernel to a set of connected
entities (stencil) is the key operation used in mesh-based
computations.

The global numbering of the degrees of freedom allows stencils to be
calculated using a direct addressing scheme when accessing the degrees
of freedom of vertically adjacent entities. We assume that the
traversal of the mesh occurs over a set of mesh entities which is
homogeneous (a set containing only cells for example). Degrees of
freedom belonging to vertically adjacent entities, accessed by two
consecutive kernel applications on the same column, have a constant
offset between them. The offset is given by the sum of degrees of
freedom attached to the two vertically adjacent entities contained in
the stencil:
\begin{equation}
\delta((d, 0)) + \delta((d, 1))
\end{equation}

Let $S$ be the stencil of a kernel which needs to access the values of
the degrees of freedom of a field $f$ defined on a function space
$\mathsf{fs}$. Let
$L_{\mathsf{fs}}(v) = (\mathsf{dof}_{0}, \mathsf{dof}_{1}, ... ,
\mathsf{dof}_{k-1})$ be the list of degrees of freedom of the stencil
for an input entity $v \in V_{d_{1}, d_{2}}$.

The lists of degrees of freedom accessed by $S$ could be provided
explicitly for all the input entities $v$. Using the previous result
we can instead reduce the number of explicitly provided lists by a
factor of $\lambda$. For each column we visit, the only explicit
accesses required are the ones to the degrees of freedom at the bottom
of the column. The degrees of freedom identifiers for the rest of the
stencil applications in the same column can be obtained by adding a
multiple of the constant vertical offset to each degree of freedom in
the bottom explicit list.

For a given stencil function $S$ an offset can be computed for each
degree of freedom in the corresponding explicit list
$L_{\mathsf{fs}}$. As the ordering of the degrees of freedom in the
stencil is fixed (by consistent ordering of mesh entities) the
vertical offset only needs to be computed once for a particular
function space $fs$.

The algorithm for computing the vertical offset is presented in
Algorithm~\ref{alg:two}. Note that since the offset for two vertically
aligned entity types is the same, only the base mesh entity type is
considered.
\begin{algorithm}[th]
\caption{Computation of vertical offsets}
\label{alg:two}
\begin{algorithmic}
\REQUIRE{$k$ : number of degrees of freedom accessed by stencil function $S$}
\REQUIRE{$E_{S}(i)$: the base mesh entity type of the $i$-th degree of freedom accessed by $S$}
\REQUIRE {$\delta((d_1, d_2))$ : the number of DoFs associated with each $(d_1, d_2)$ entity}
\ENSURE{$\mathsf{offset}_{S, \mathsf{fs}}$ : the vertical offset for function space $\mathsf{fs}$ given stencil $S$}

\FOR{$i$ in $\{0, 1, ..., k-1\}$}
	\STATE $d \leftarrow E_{S}(i)$
	\STATE $\mathsf{offset}_{S, \mathsf{fs}}(i) \leftarrow \delta((d, 0)) + \delta((d, 1))$
\ENDFOR
\end{algorithmic}
\end{algorithm}

If $(\mathsf{dof}_{0}, \mathsf{dof}_{1}, ... , \mathsf{dof}_{k-1})$ is
the explicit list of degrees of freedom for the initial layer to which
the stencil can be applied, then the list of degrees of freedom for
the $n^\mathrm{th}$ application of the stencil along the vertical is
given by:
\begin{equation}
(\mathsf{dof}_{0} + n \times (\mathsf{offset}_{S, f}(0)) , ... , \mathsf{dof}_{k-1} + n \times (\mathsf{offset}_{S, f}(k-1)))
\end{equation}

Algorithm~\ref{alg:three} shows the iteration algorithm working for a
single field $f$ on a function space $\mathsf{fs}$. The stencil
function $S$ is applied to the entities of each column in turn. Each
time the algorithm moves on to the next vertically adjacent entity,
the indices of the degrees of freedom accessed are incremented by the
vertical offset $\mathsf{offset}_{S, \mathsf{fs}}$. The algorithm is
also applicable to stencil functions of multiple fields defined on the
same function space since the data associated with each field is
accessible using the same set of degree of freedom numbers.  The
extension to fields from different function spaces just requires
explicit lists $L_{\mathsf{fs}}$ for each space.

\begin{algorithm}[th]
\caption{Iteration of a stencil function over an extruded mesh}
\label{alg:three}
\begin{algorithmic}
\REQUIRE{$V$: iteration set of base mesh entities}
\REQUIRE {$d_2$ : the dimension of vertical iteration entities}
\REQUIRE {$\lambda$ : the number of vertical intervals}
\REQUIRE{$S$: stencil function to be applied to the degrees of freedom of field $f$}
\REQUIRE{$L_{\mathsf{fs}}$: set of explicit lists of degrees of freedom for function space $\mathsf{fs}$}
\REQUIRE{$\mathsf{offset}_{S, \mathsf{fs}}$: the vertical offset for function space $\mathsf{fs}$ given stencil $S$}
\FOR{$v$ in $V$}
		\STATE $(\mathsf{dof}_{0}, \mathsf{dof}_{1}, ..., \mathsf{dof}_{k-1}) \leftarrow L_{\mathsf{fs}}(v)$
		\FOR {$l$ in $\{0, 1, ... , \lambda - d_{2}\}$}
			\STATE $S(f(\mathsf{dof}_{0}), f(\mathsf{dof}_{1}), ..., f(\mathsf{dof}_{k-1}))$
			\FOR {$j$ in $\{0, 1, ..., k-1\}$}
				\STATE $\mathsf{dof}_{j} \leftarrow \mathsf{dof}_{j}+\mathsf{offset}_{S,\mathsf{fs}}(j)$
			\ENDFOR
		\ENDFOR
\ENDFOR
\end{algorithmic}
\end{algorithm}

\section{Performance Evaluation}
\label{sec:performance-evaluation}

In this section, we test the hypothesis that iteration exploiting the
extruded structure of the mesh amortizes the unstructured base mesh
overhead of accessing memory through explicit neighbour lists. We also
show that the more layers the mesh contains, the closer its
performance is to the hardware limits of the machine.

We validate our hypotheses in the Firedrake finite element framework
\citep{Rathgeber:2016}. Although we restrict our performance evaluation
to examples drawn from finite element discretisations, the algorithms
we have presented can be applied to any mesh-based discretisation.

In \autoref{ssec:experimental-design} we describe the design of the
experiments undertaken. The hardware platforms and the methodology
used are described in \autoref{ssec:experimental-setup} followed by
results and discussion in \autoref{ssec:experimental-results} and
\autoref{ssec:discussion} respectively.

\subsection{Experimental Design}
\label{ssec:experimental-design}

The design space to be explored is parameterized by number of layers
and the manner in which the data is associated with the mesh and
therefore accessed. In establishing the relationship between the
performance and the hardware we examine performance on two generations
of processors and varying process counts.

\subsubsection{Choosing the computation}
\label{sssec:choosing-computation}
Numerical computations of integrals are the core mesh iteration
operation in the finite element method. We focus on residual (vector)
assembly for two reasons. First, in contrast to Jacobian assembly,
there are no overheads due to sparse matrix insertion; the experiment
is purely a test of data access via the mesh indirections. Second,
residual evaluation is the assembly operation with the lowest
computational intensity and therefore constitutes a worst-case
scenario for data layout performance exploration.

Since we are interested in data accesses, we choose the simplest
non-trivial residual assembly operation:
\begin{equation}
I_{1} = \int_{\Omega}\! f v \,\mathrm{d}x, \quad \forall v \in V
\end{equation}
for $f$ in the finite element space $V$.  For this study we choose
$\Omega = [0, 1]^3$ to be the unit cube.  The base mesh is generated
in an unstructured manner using Gmsh \citep{Geuzaine:2009}, and then
extruded to form a three-dimensional domain.

In addition to the output field $I_{1}$ and the input field $f$ this
computation accesses the coordinate field, $\vec{x}$. Regardless of
the choice of $V$, we always represent $\vec{x}$ by a $d$-vector at
each vertex of the $d$-dimensional mesh.

\subsubsection{Choosing the discretisations}
\label{sssec:choosing-discretisation}

The construction of a wide variety of finite element spaces on
extruded meshes was introduced in \citet{McRae:2016}. This enables us
to select the horizontal and vertical data discretisations
independently.

For the purposes of data access, the distinguishing feature of
different finite element spaces is the extent to which degrees of
freedom are shared between adjacent cells.

We choose a set of finite element spaces spanning the combinations of
horizontal and vertical reuse patterns found on extruded meshes:
horizontal and vertical reuse, only horizontal, only vertical, or no
reuse at all.

We employ low order continuous and discontinuous discretisations
(abbreviated as \emph{CG} and \emph{DG} respectively) in both the
horizontal and vertical directions.

\begin{figure}
  \centering
  \subcaptionbox{$\mathrm{CG1} \times \mathrm{CG1}$, horizontal and
    vertical reuse}[0.3\linewidth]
  {\includegraphics[width=0.25\linewidth]{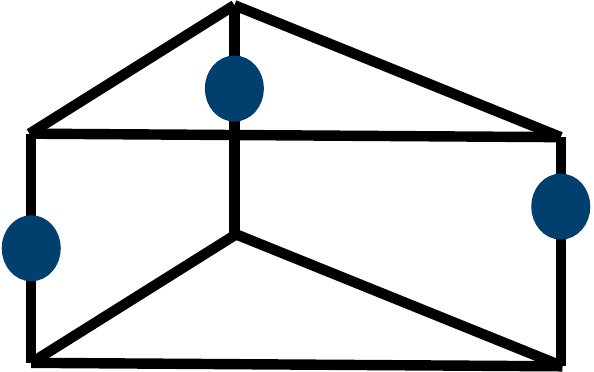}}
  \subcaptionbox{$\mathrm{CG1} \times \mathrm{DG0}$, horizontal reuse}
  [0.3\linewidth]
  {\includegraphics[width=.25\linewidth]{assets/PrismCG1xDG0}}
  \subcaptionbox{$\mathrm{CG1} \times \mathrm{DG1}$, horizontal reuse}
  [0.3\linewidth]
  {\includegraphics[width=.25\linewidth]{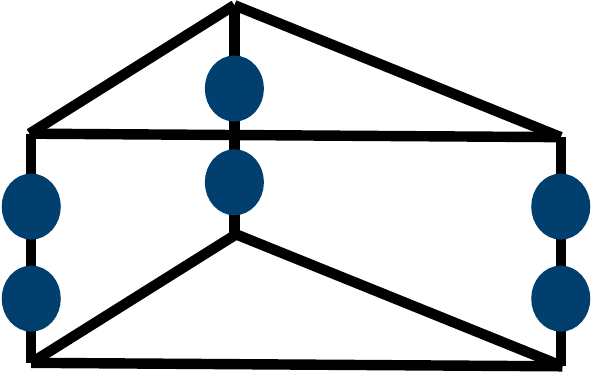}}

  \vspace{1em}
  \subcaptionbox{$\mathrm{DG0} \times \mathrm{CG1}$, vertical reuse}
  [0.3\linewidth]
  {\includegraphics[width=.25\linewidth]{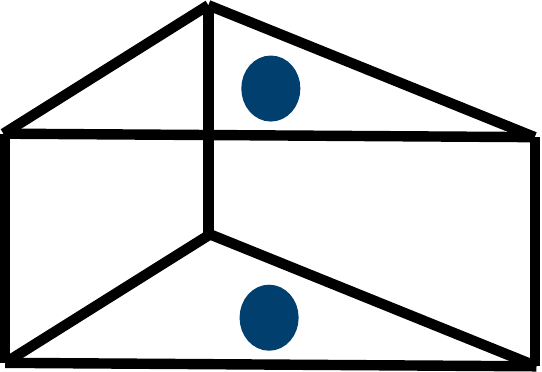}}
  \subcaptionbox{$\mathrm{DG0} \times \mathrm{DG0}$, no reuse}
  [0.3\linewidth]
  {\includegraphics[width=.25\linewidth]{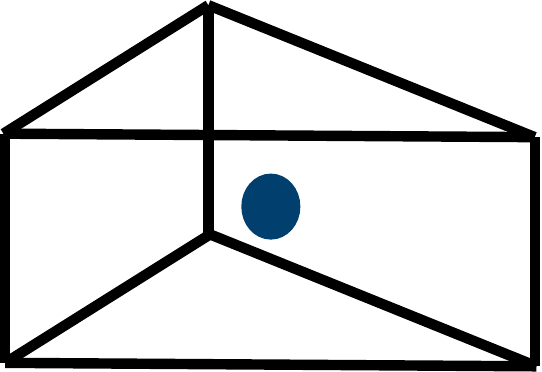}}
  \subcaptionbox{$\mathrm{DG0} \times \mathrm{DG1}$, no reuse}
  [0.3\linewidth]
  {\includegraphics[width=.25\linewidth]{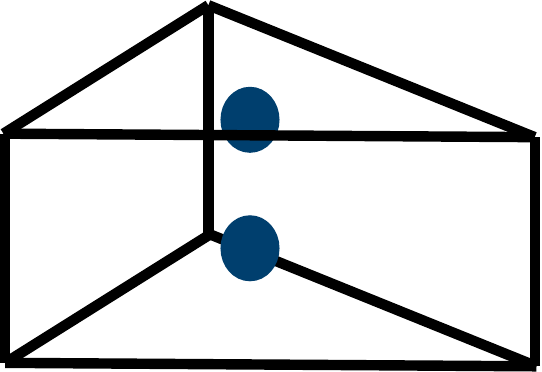}}

  \vspace{1em}
  \subcaptionbox{$\mathrm{DG1} \times \mathrm{CG1}$, vertical reuse}
  [0.3\linewidth]
  {\includegraphics[width=.25\linewidth]{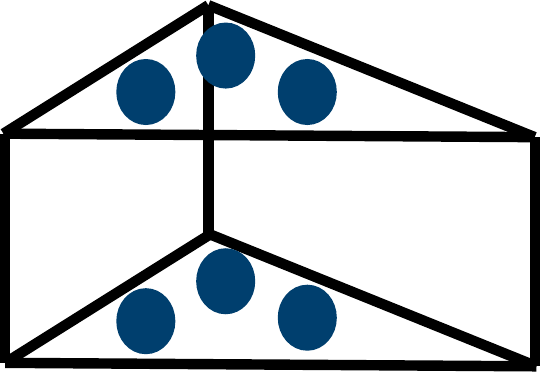}}
  \subcaptionbox{$\mathrm{DG1} \times \mathrm{DG0}$, no reuse}
  [0.3\linewidth]
  {\includegraphics[width=.25\linewidth]{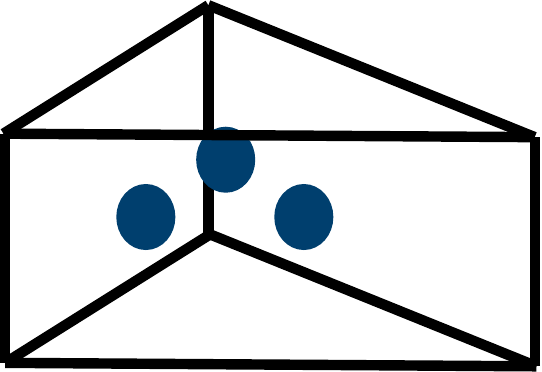}}
  \subcaptionbox{$\mathrm{DG1} \times \mathrm{DG1}$, no reuse}
  [0.3\linewidth]
  {\includegraphics[width=.25\linewidth]{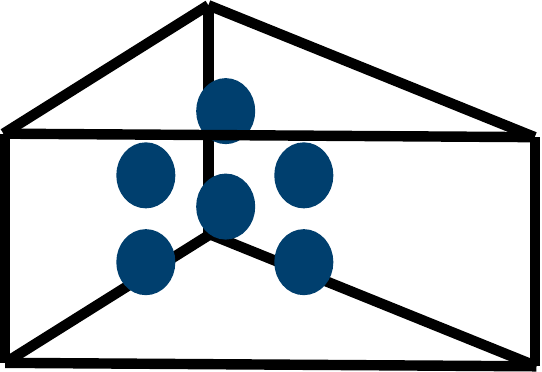}}
  \caption{Tensor product finite elements with different data layout
    and cell-to-cell data re-use.}
\label{fig:tpe-data-layout}
\end{figure}

The set of discretisations is
$A = \{\mathrm{CG1}, \mathrm{DG0}, \mathrm{DG1}\}$ where the number
indicates the degree of polynomials in the space. We examine all pairs
of discretisations $ (h , v) \in A \times A$. Since the cells of the
base mesh are triangles, the extruded mesh consists of triangular
prisms. \autoref{fig:tpe-data-layout} shows the data layout of each of
these finite elements.

Both Firedrake and our numbering algorithm support a much larger range
of finite element spaces than this. However, the more complex and
higher degree spaces will result in more computationally intensive
kernels but not materially different data reuse. The lowest order
spaces are the most severe test of our approach since they are more
likely to be memory bound.

\subsubsection{Layer count and problem size}
\label{sssec:layer-count}
We vary the number of layers between 1 and 100. This is a realistic
range for current ocean and atmosphere simulations. The number of
cells in the extruded mesh is kept approximately constant by shrinking
the base mesh as the number of layers increases. The mesh size is
chosen such that the data volume far exceeds the total last level
cache capacity of each chosen architecture (L3 cache in all
cases). This minimizes caching benefits and is therefore the strongest
test of our algorithms. The overall mesh size is fixed at
approximately 15 million cells which yields a data volume of between
300 and 840 MB depending on discretisation.

\subsubsection{Base mesh numbering}
\label{sssec:base-numbering}
The order in which the entities of the unstructured mesh are numbered
is known to be critical for data access performance. To characterize
this effect and distinguish it from the impact of the number of
layers, we employ two variants of each base mesh. The first is a mesh
for which the traversal is optimised using a reverse Cuthill-McKee
ordering \citep{Cuthill:1969}. The second is a \emph{badly} ordered
mesh with a random numbering. This represents a pathological case for
temporal locality.

\subsection{Experimental Setup}
\label{ssec:experimental-setup}

The specification of the hardware used to conduct the experiments is
shown in \autoref{tab:hardware}. Following \citet{Ofenbeck:2014} we
disable the Intel turbo boost and frequency scaling. This is intended
to prevent our performance results from being subject to fluctuations
due to processor temperature.
\begin{table}[htbp]
  \centering
  \begin{tabular}{l|l|l}
    Name                 & Intel Sandy Bridge & Intel Haswell   \\
    \hline
    Model                & Xeon E5-2620       & Xeon E5-2640 v3 \\
    Frequency            & 2.0 GHz            & 2.6 GHz         \\
    Sockets              & 2                  & 2               \\
    Cores per socket     & 6                  & 8               \\
    Bandwidth per socket & 42.6 GB/s          & 56.0 GB/s       \\
  \end{tabular}
  \caption{Hardware used.\label{tab:hardware}}
\end{table}%

The experiments we are considering are run on a single two-socket
machine and use MPI (Message Passing Interface) parallelism. The
number of MPI processes varies from one up to 2 processes per physical
core (exploiting hyperthreading). We pin the processes evenly across
physical cores to ensure load balance and prevent process migration
between cores.

The Firedrake platform performs integral computations by automatically
generating \emph{C} code. The compiler used is GCC version 4.9.1
(\texttt{-march=native -fassociative-math -ffast-math -O3}).  We also
assessed the performance of the Intel C Compiler version 15.0.2
(\texttt{-O3 -xAVX -ip -xHost}), however we only report results from
GCC in this paper since the performance of the Intel compiler was
inferior.

\subsubsection{Runtime, data volume, bandwidth and FLOPs}
\label{sssec:runtime}
Runtime is measured using a nanosecond precision timer. Each
experiment is performed ten times and we report the minimum
runtime. Exclusive access to the hardware has been ensured for all
experiments.

We model the data transfer from main memory to CPU assuming a perfect
cache: each piece of data is only loaded from main memory once. We
define the \emph{valuable data volume} as the total size of the input,
output and coordinate fields. This gives a lower bound on the memory
traffic to and from main memory. The valuable data volume divided by
the runtime yields the \emph{valuable bandwidth}.

Different discretisations lead to different data volumes due to the
way data is shared between cells. $\mathrm{DG}$ based discretisations
require the movement of larger data volumes while $\mathrm{CG}$
discretisations lead to smaller volumes due to data reuse.

To evaluate the impact of different data volumes we compare the
valuable bandwidth with the maximum bandwidth achieved for the STREAM
triad benchmark~\citep{McCalpin:1995}, shown in
\autoref{tab:stream-triad}.  The valuable bandwidth achieved as
percentage of STREAM bandwidth shows how prone the code is to becoming
bandwidth bound as its floating point performance is improved.
\begin{table}[htbp]
  \centering
  \begin{tabular}{l|c}
    Platform           & STREAM bandwidth \\
    \hline
    Intel Sandy Bridge & 55.3 GB/s        \\
    Intel Haswell      & 80.2 GB/s        \\
  \end{tabular}
  \caption{Maximum STREAM triad ($a_i = b_i + \alpha c_i$) performance
    achieved by varying the number of MPI processes from one to twice
    the number of physical cores.\label{tab:stream-triad}}
\end{table}

The floating point operations -- adds, multiplies and, on Haswell,
fused multiply-add (FMA) operations -- are counted automatically using
the Intel Architecture Code Analyzer~\citep{iaca} whose results are
verified with PAPI~\citep{Mucci:1999} which accesses the hardware
counters.

\subsubsection{Theoretical performance bounds}
\label{sssec:performance-bounds}

The performance of the extruded iteration depends on the efficiency of
the generated finite element kernel (payload) code which for some
cases may not be vectorised (as outlined in \citet{Luporini:2015}) or
may not have a perfectly balanced number of floating point additions
and multiplications. A discussion of kernel code optimality is outside
the scope of this paper.

To a first approximation the performance of a numerical algorithm will
be limited by either the memory bandwidth or the floating point
throughput. The STREAM benchmark provides an effective upper bound on
the achievable memory bandwidth. The floating point bounds employed
are based on the theoretical maximum given the clock frequency of the
processor.

The Intel architectures considered are capable of executing both a
floating point addition and a floating point multiplication on each
clock cycle. The Haswell processor can execute a fused multiply-add
instruction (FMA) instead of either an addition or multiplication
operation.

The achievable FLOP rate may therefore be as much as twice the clock
rate depending on the mix of instructions executed. The achievable
speed-up over the clock rate, $f_{b}$, for the Sandy Bridge platform
is therefore bounded by the balance factor
\begin{equation}
f_{b} = 1 + \frac{\min(\mathrm{add\ FLOPs}, \mathrm{multiplication\ FLOPs})}  {\max(\mathrm{add\ FLOPs}, \mathrm{multiplication\ FLOPs})},
\end{equation}
while for Haswell it is bounded by
\begin{equation}
f_{b} = 1 + \frac{\min(\mathrm{add\ FLOPs}, \mathrm{multiplication\ FLOPs}) + k}  {\max(\mathrm{add\ FLOPs}, \mathrm{multiplication\ FLOPs}) + k},
\end{equation}
where $k$ is half the number of FMAs.

\subsubsection{Vectorisation}
\label{sssec:vectorisation}

The processors employed support 256-bit wide vector floating point
instructions. The double precision FLOP rate of a fully vectorised
code can be as much as four times that of an unvectorised code. GCC
automatically vectorised only a part of the total number of floating
point instructions. The ratio between the number of vector (packed)
floating point instructions and the total number of floating point
instructions (scalar and packed) characterizes the impact of partial
vectorization on the floating point bound through the vectorization
factor
\begin{equation}
f_{v} = 1 + (4 - 1) \times \frac{\mathrm{vector\ FLOPs}} {\mathrm{total\ FLOPs}}.
\end{equation}

To control the impact of the kernel computation (payload) on the
evaluation, we compare the measured floating point throughput with a
theoretical peak which incorporates the payload instruction balance
and the degree of vectorization. Let $c$ be the number of active
physical CPU cores during the computation of interest. The theoretical
base floating point performance $B_{c}$ is the same for all
discretisations and assumes one floating point instruction per cycle
for each active physical CPU core. The peak theoretical floating point
throughput $P_{d}$ is different for each discretisation $d$ as it
depends on the properties of the payload and is given by
\begin{equation}
P_{d} = B_{c} \times f_{b} \times f_{v}.
\end{equation}

\subsection{Experimental Results}
\label{ssec:experimental-results}

\subsubsection{Percentage of theoretical performance}
\label{sssec:percentage-theoretical}
For the Sandy Bridge and Haswell architectures, the best performance
is achieved in the 100-layer case run with 24 and 32 processes
respectively (hyperthreading enabled).
The results in \autoref{tab:sandy-bridge-percentage-peak} and
\autoref{tab:haswell-percentage-peak} show percentages of the STREAM
bandwidth and the theoretical floating point throughput which
incorporates the instruction balance and vectorization factors.
\begin{table}[htbp]
\centering
\begin{tabular}{l|l|l|c|c}
Discretisation                     & $f_{b}$ & $f_{v}$ & $P_{d}$ (\%) & Bandwidth (\%) \\
\hline
$\mathrm{CG1} \times \mathrm{CG1}$ & 1.7     & 1.58    & 73.45        & 7.092          \\
$\mathrm{CG1} \times \mathrm{DG0}$ & 1.81    & 1.0     & 78.96        & 14.70          \\
$\mathrm{CG1} \times \mathrm{DG1}$ & 1.7     & 1.58    & 73.03        & 10.50          \\
$\mathrm{DG0} \times \mathrm{CG1}$ & 1.65    & 1.0     & 76.01        & 27.86          \\
$\mathrm{DG0} \times \mathrm{DG0}$ & 1.5     & 1.0     & 85.14        & 34.86          \\
$\mathrm{DG0} \times \mathrm{DG1}$ & 1.65    & 1.0     & 75.45        & 45.68          \\
$\mathrm{DG1} \times \mathrm{CG1}$ & 1.7     & 1.58    & 73.20        & 24.60          \\
$\mathrm{DG1} \times \mathrm{DG0}$ & 1.81    & 1.0     & 78.93        & 50.98          \\
$\mathrm{DG1} \times \mathrm{DG1}$ & 1.7     & 1.58    & 71.78        & 44.37          \\
\end{tabular}
\caption{Percentage of STREAM bandwidth and theoretical throughput
  achieved by the computation of integral $I$ over 100 layers on Sandy
  Bridge with 24 MPI
  processes.\label{tab:sandy-bridge-percentage-peak}}
\end{table}
 
\begin{table}[htbp]
  \centering
  \begin{tabular}{l|l|l|c|c}
    Discretisation                      & $f_{b}$ & $f_{v}$ & $P_{d}$ (\%) & Bandwidth (\%) \\
    \hline
    $\mathrm{CG1} \times \mathrm{CG1}$ & 1.76    & 1.61    & 72.43        & 9.015          \\
    $\mathrm{CG1} \times \mathrm{DG0}$ & 1.97    & 1.0     & 88.57        & 21.92          \\
    $\mathrm{CG1} \times \mathrm{DG1}$ & 1.76    & 1.61    & 72.20        & 13.39          \\
    $\mathrm{DG0} \times \mathrm{CG1}$ & 1.87    & 1.0     & 73.94        & 38.74          \\
    $\mathrm{DG0} \times \mathrm{DG0}$ & 1.66    & 1.0     & 91.93        & 53.10          \\
    $\mathrm{DG0} \times \mathrm{DG1}$ & 1.87    & 1.0     & 72.89        & 63.11          \\
    $\mathrm{DG1} \times \mathrm{CG1}$ & 1.76    & 1.61    & 71.99        & 31.19          \\
    $\mathrm{DG1} \times \mathrm{DG0}$ & 1.97    & 1.0     & 87.55        & 75.17          \\
    $\mathrm{DG1} \times \mathrm{DG1}$ & 1.76    & 1.61    & 71.50        & 56.98          \\
  \end{tabular}
  \caption{Percentage of STREAM bandwidth and theoretical throughput
    achieved by the computation of integral $I$ over 100 layers on
    Haswell with 32 MPI processes.\label{tab:haswell-percentage-peak}}
\end{table}

On Sandy Bridge, the proportion of peak theoretical floating point
throughput is between 71 and 85\%, while on Haswell it is between 71
and 92\%.  In contrast, the proportion of peak bandwidth achieved
varies between 7 and 51\% on Sandy Bridge and 9 and 75\% on Haswell.
The higher, and much more consistent peak FLOP results lead us to the
conclusion that we are in an operation- rather than bandwidth-limited
regime.  The performance figures are therefore presented with respect
to this metric.

\subsubsection{Amortizing the cost of indirect accesses}
\label{sssec:amortizing-indirect}

\begin{figure}[p]
\centering
  \subcaptionbox{Sandy Bridge, 1 process, $c=1$}
  [0.48\linewidth]
  {\includegraphics[width=.45\linewidth]{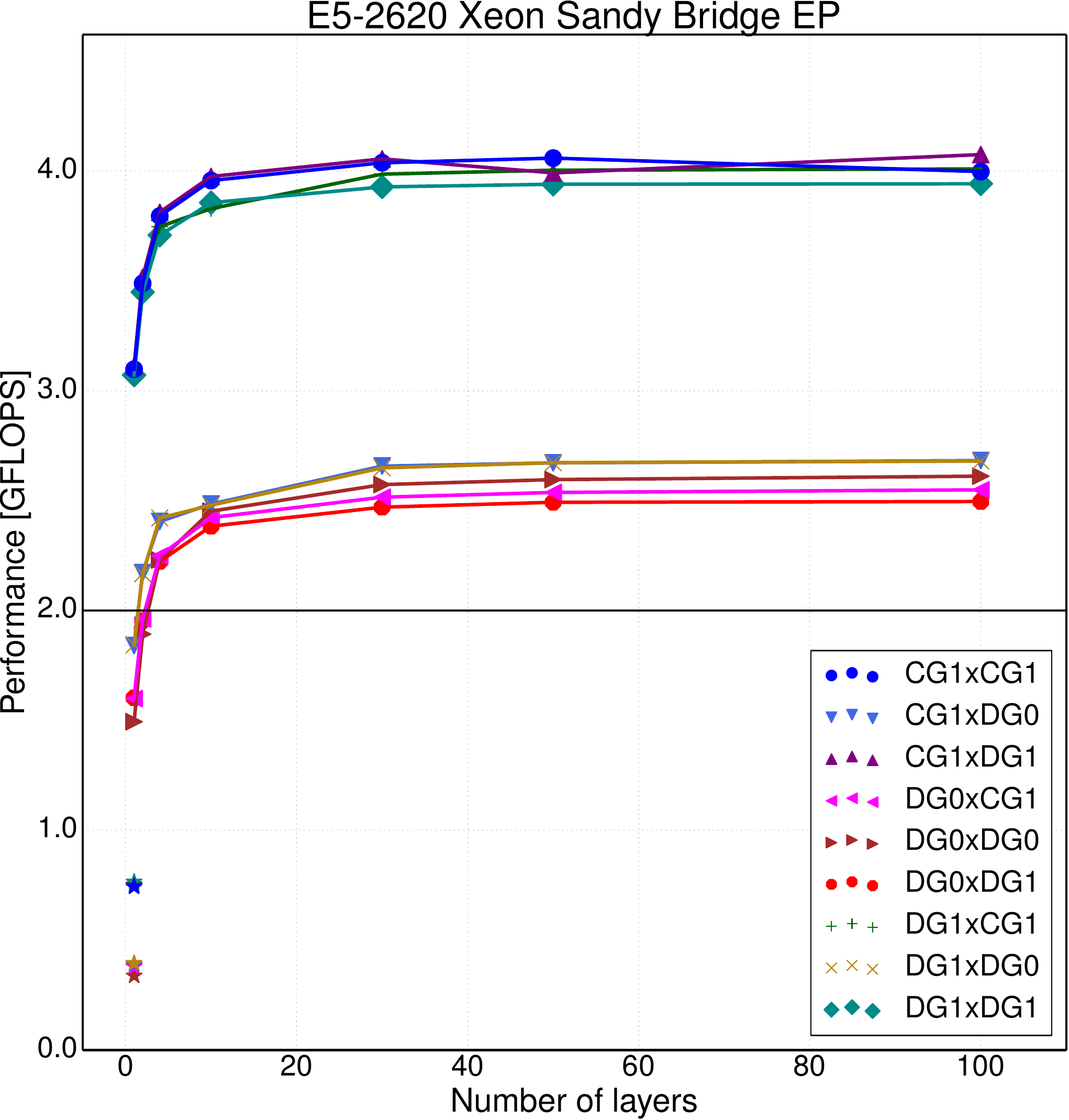}}
  \subcaptionbox{Sandy Bridge, 6 processes, $c=6$}
  [0.48\linewidth]
  {\includegraphics[width=.45\linewidth]{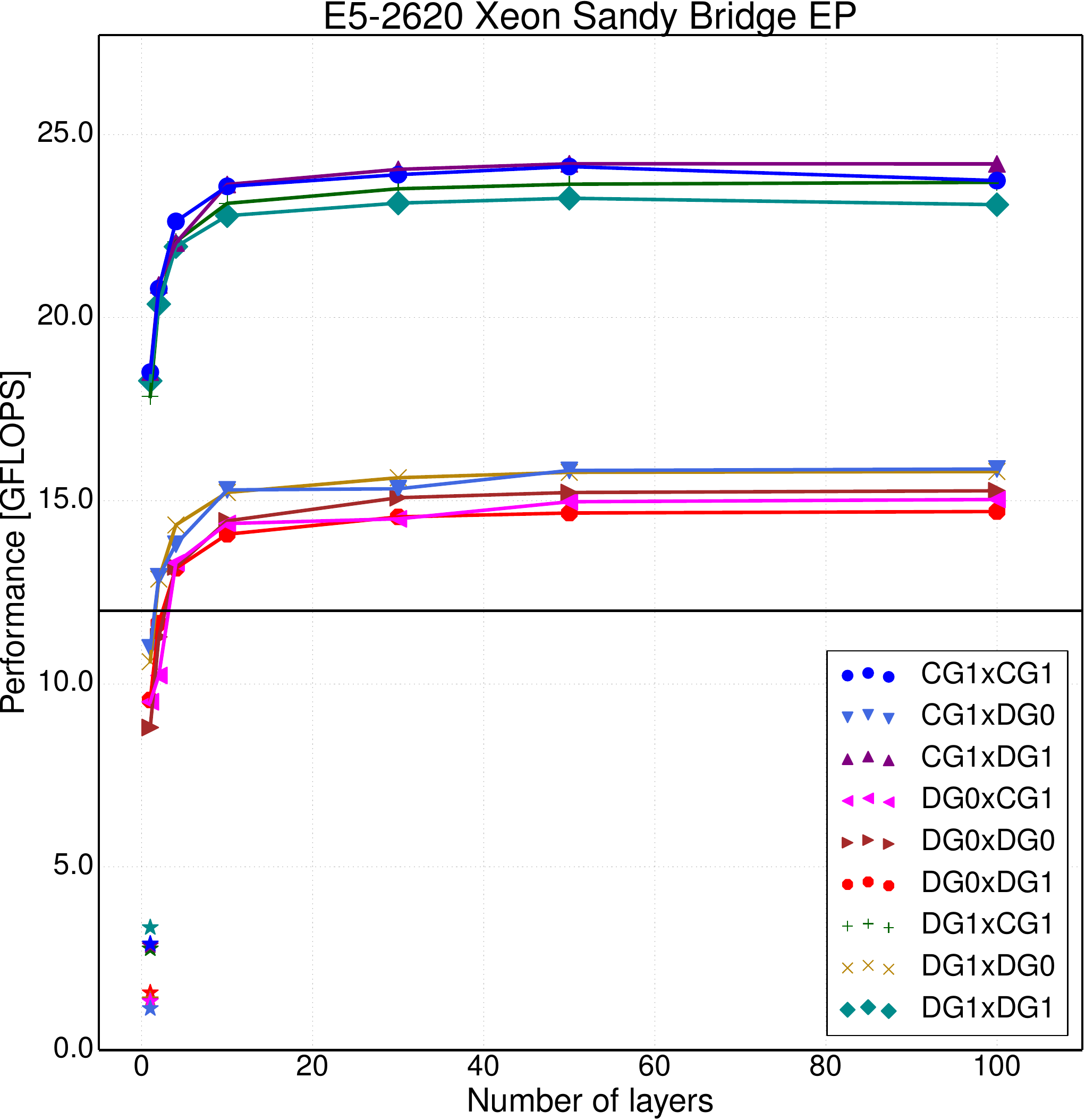}}

  \vspace{\baselineskip}
  \subcaptionbox{Sandy Bridge, 12 processes, $c=12$}
  [0.48\linewidth]
  {\includegraphics[width=.45\linewidth]{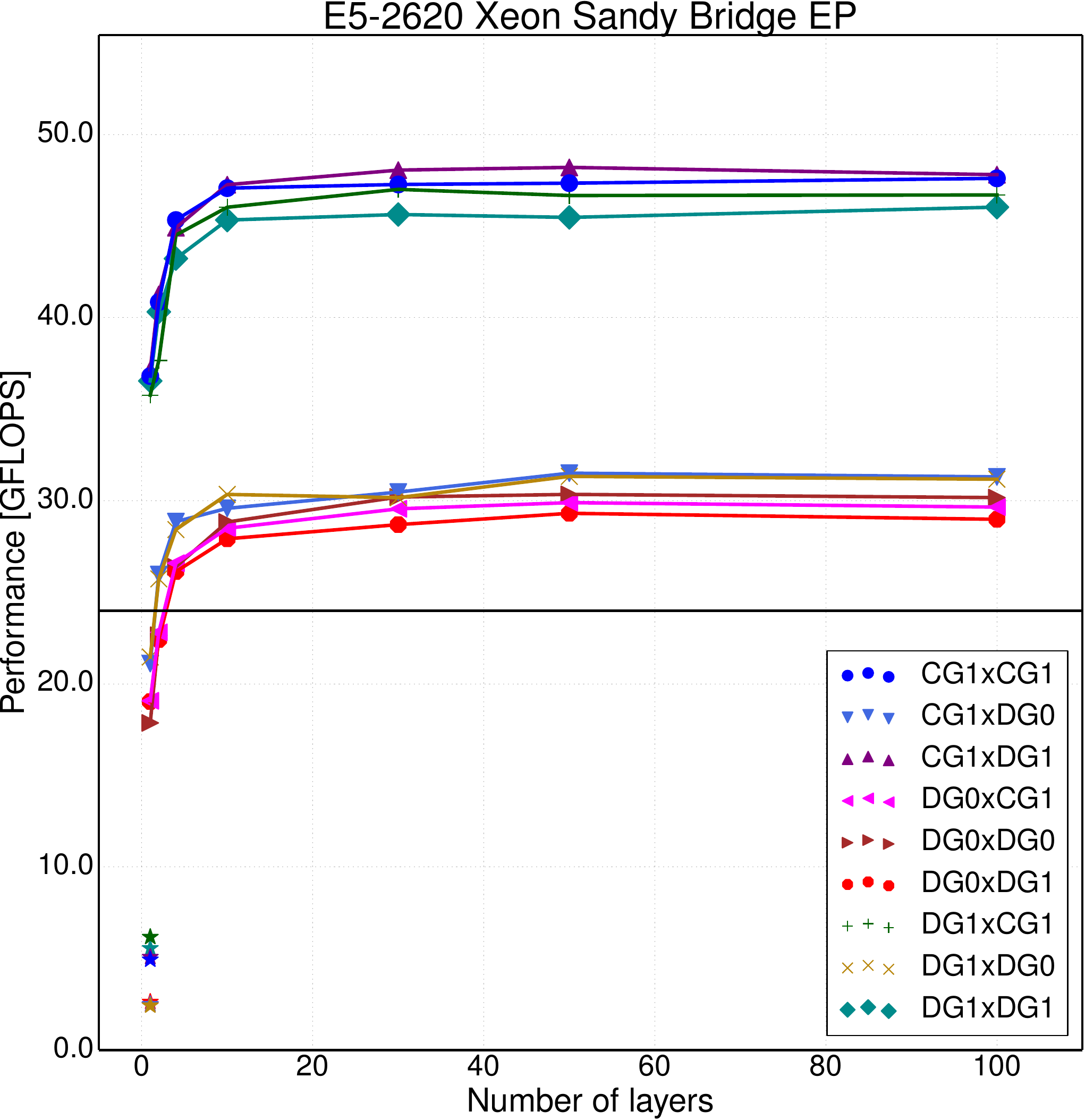}}
  \subcaptionbox{Sandy Bridge, 24 processes, $c=12$}
  [0.48\linewidth]  
  {\includegraphics[width=.45\linewidth]{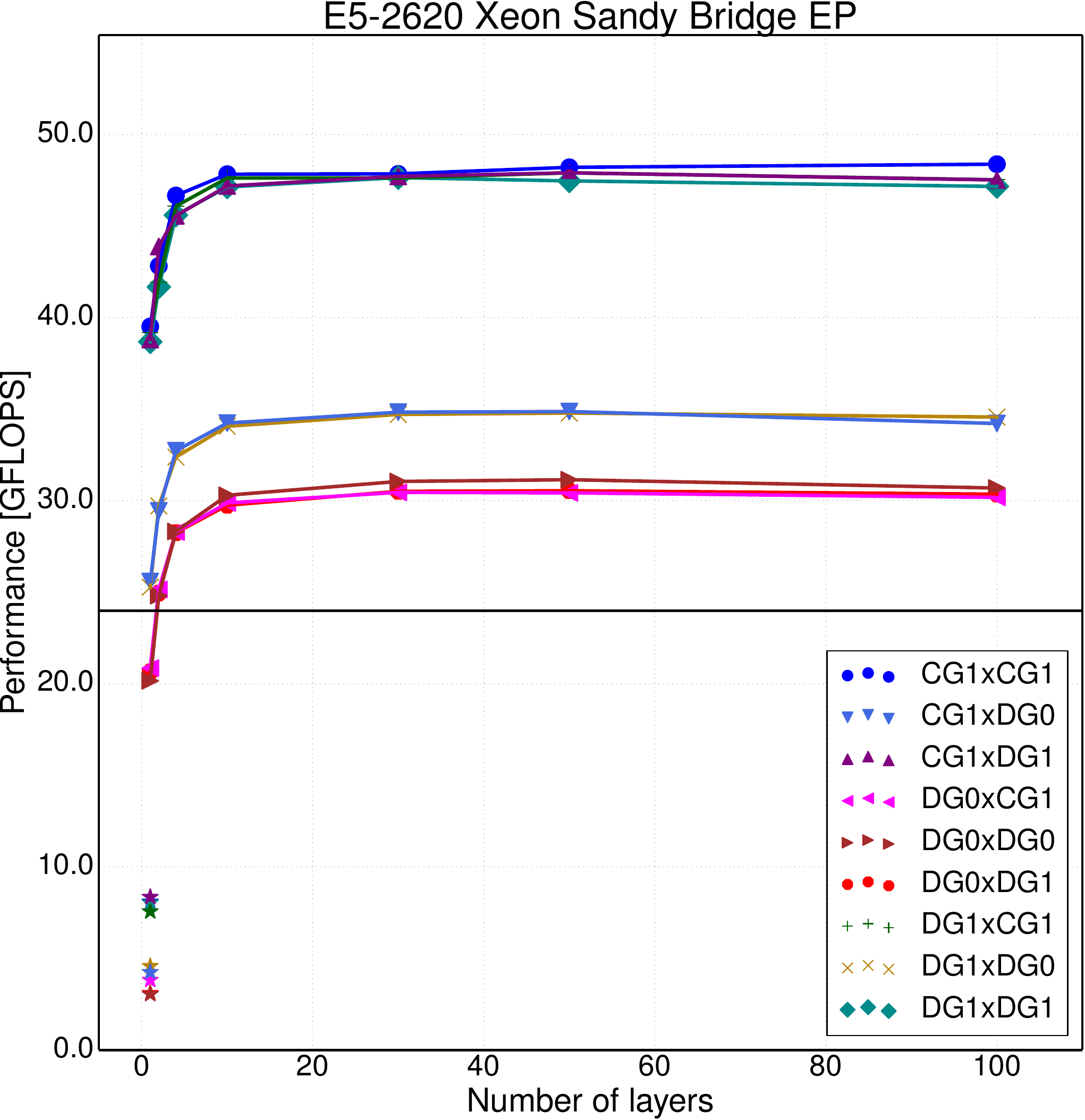}}
  \caption{Performance of the $I$ integral computation with varying
    number of layers and number of processes on a well-ordered base
    mesh. The star-shaped markers show the performance of the 1-layer
    badly-ordered mesh for comparison. The horizontal line is the base
    FLOP throughput for $f_b = f_v = 1$ and the number of physical
    cores used.}
  \label{fig:sandy-bridge-good-ordering}
\end{figure}

\begin{figure}[p]
  \centering
  \subcaptionbox{Sandy Bridge, 1 process, $c=1$}
  [0.48\linewidth]
  {\includegraphics[width=.45\linewidth]{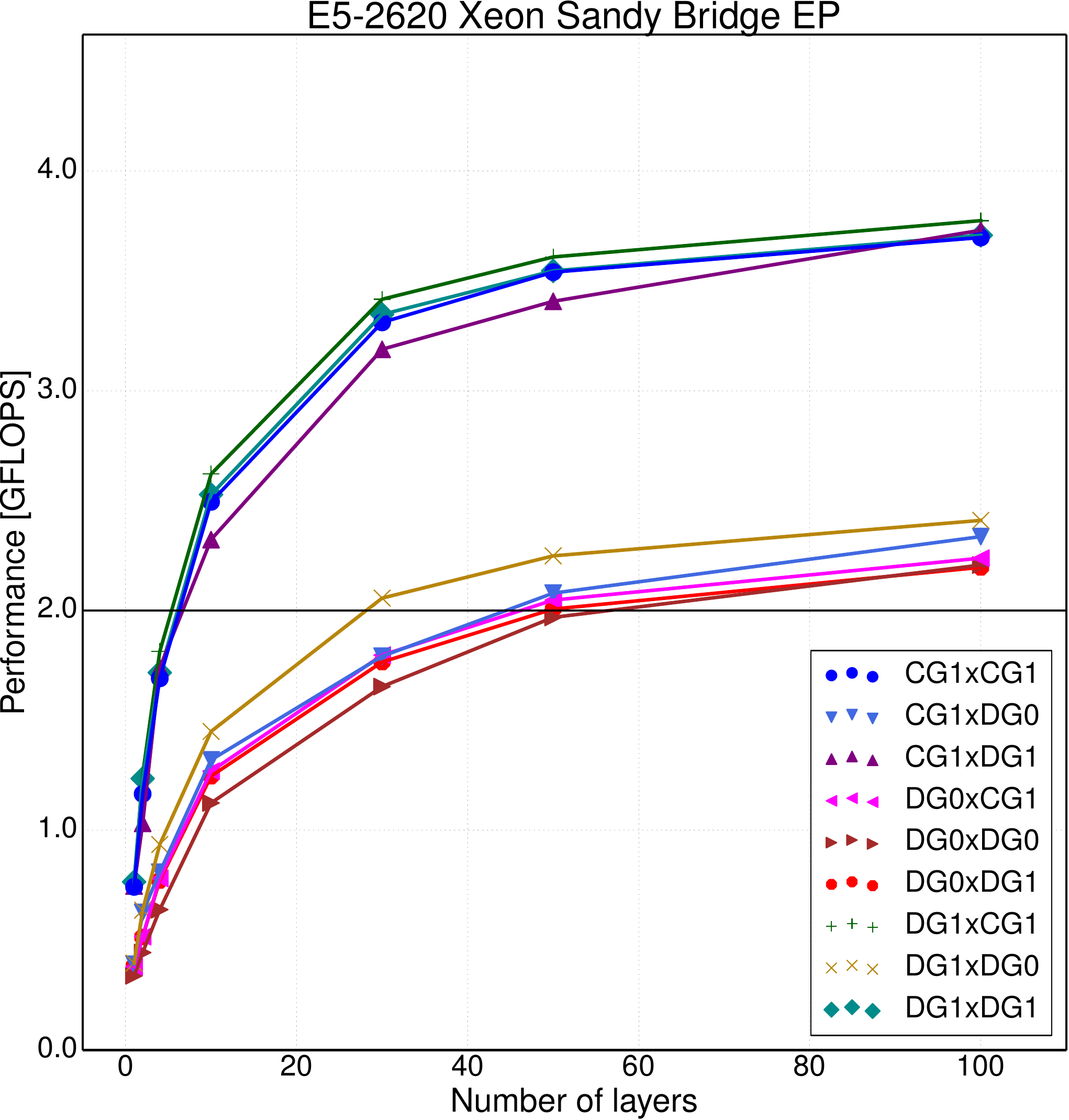}}
  \subcaptionbox{Sandy Bridge, 6 processes, $c=6$}
  [0.48\linewidth]
  {\includegraphics[width=.45\linewidth]{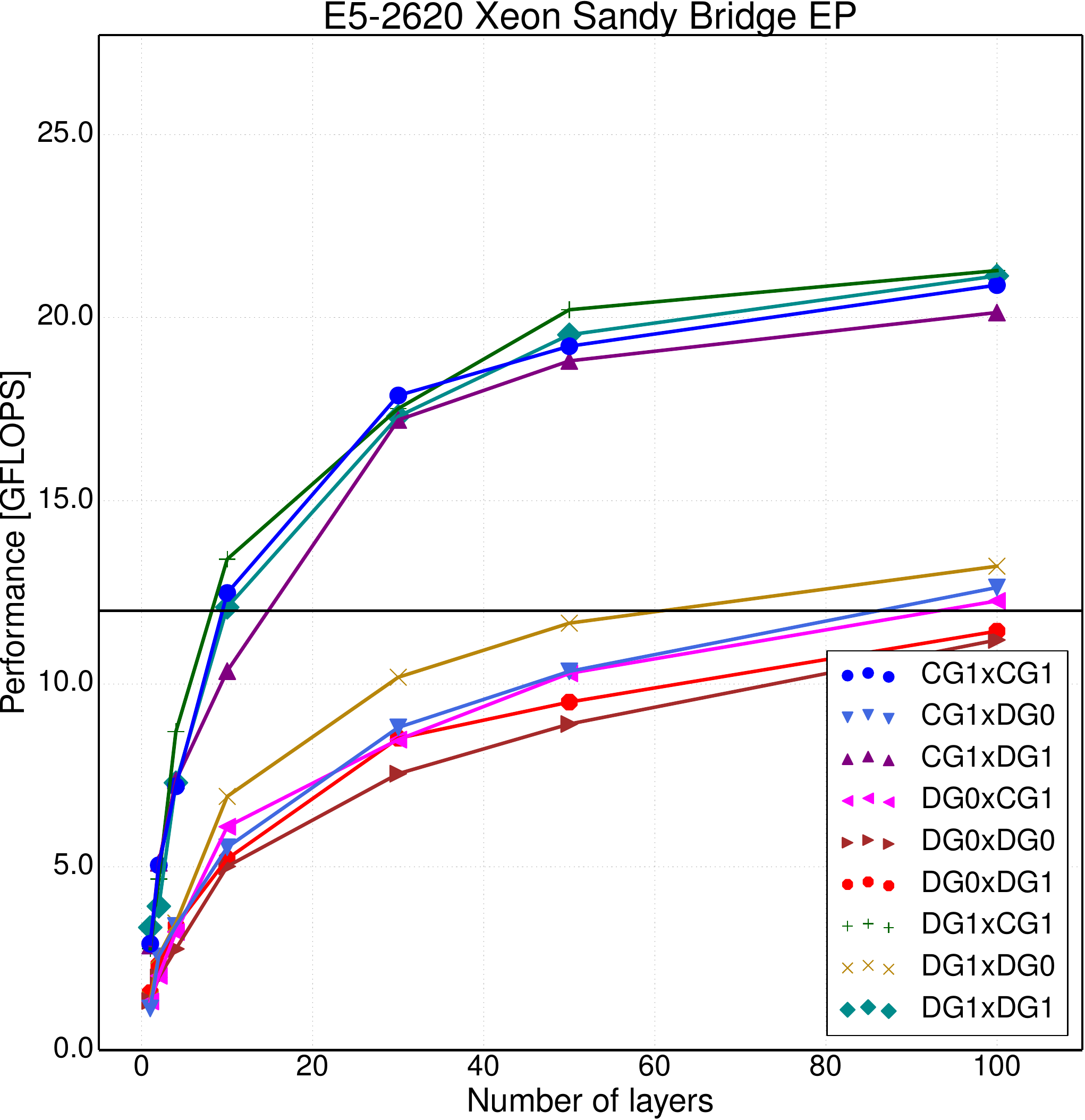}}

  \vspace{\baselineskip}
  \subcaptionbox{Sandy Bridge, 12 processes, $c=12$}
  [0.48\linewidth]
  {\includegraphics[width=.45\linewidth]{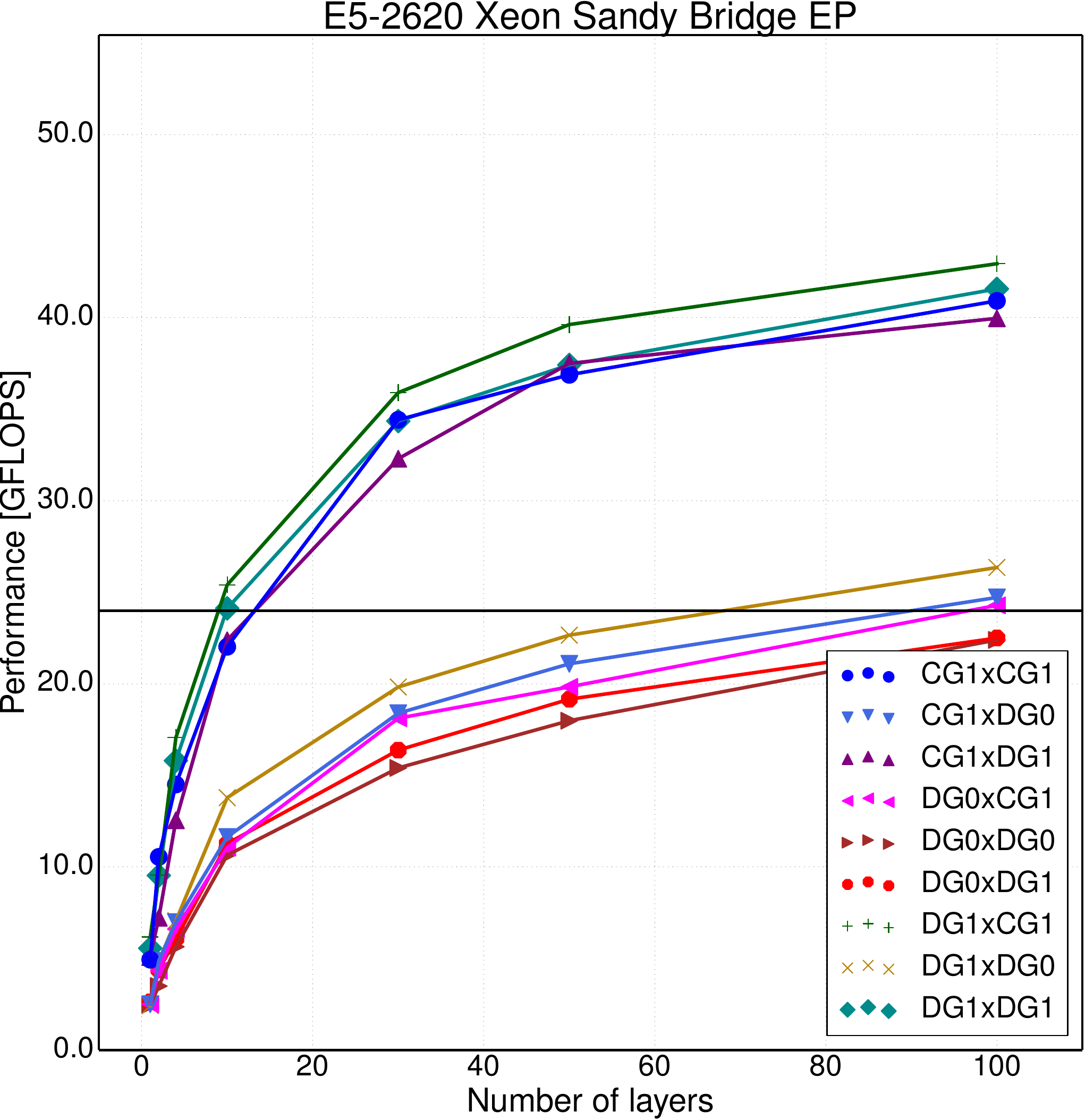}}
  \subcaptionbox{Sandy Bridge, 24 processes, $c=12$}
  [0.48\linewidth]  
  {\includegraphics[width=.45\linewidth]{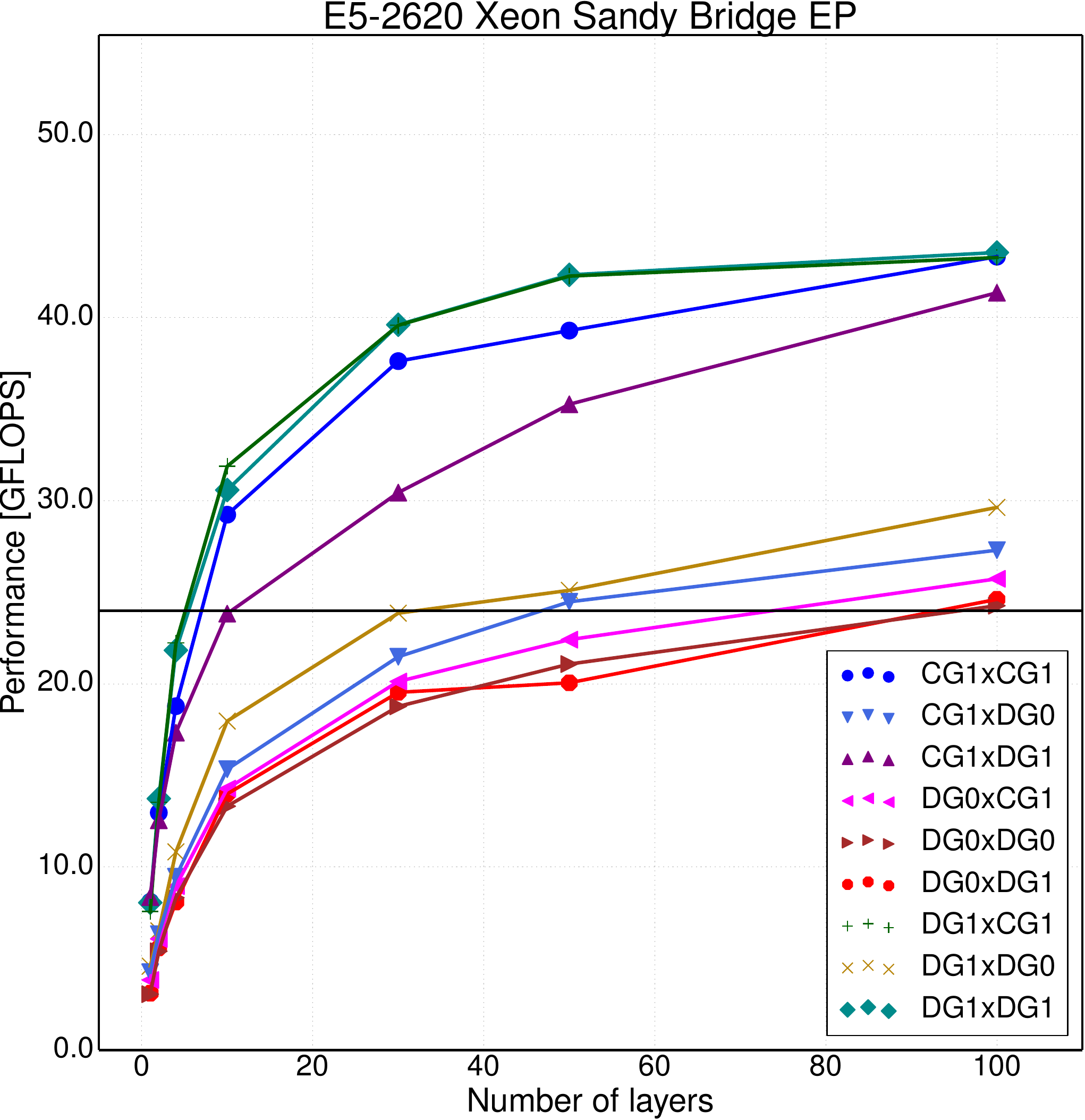}}

  \caption{Performance of the $I$ integral computation with varying
    number of layers and number of processes on a badly-ordered base
    mesh. The horizontal line is the base FLOP throughput for
    $f_b = f_v = 1$ and the number of physical cores used.}
  \label{fig:sandy-bridge-bad-ordering}
\end{figure}

When the base mesh is well ordered
(\autoref{fig:sandy-bridge-good-ordering}), the number of layers
required to reach a performance plateau is between 10 and 20 for all
discretisations. When the base mesh is badly ordered
(\autoref{fig:sandy-bridge-bad-ordering}) the plateau is frequently
not reached even with 100 layers.  A striking feature of both
\autoref{fig:sandy-bridge-good-ordering} and
\autoref{fig:sandy-bridge-bad-ordering} is that cases in which the
local kernel calculations are identical produce very similar achieved
FLOP rates, despite having different data sharing patterns.  This
supports the hypothesis that the results are operation bound.

\begin{figure}
  \centering
  \includegraphics[width=.5\linewidth]{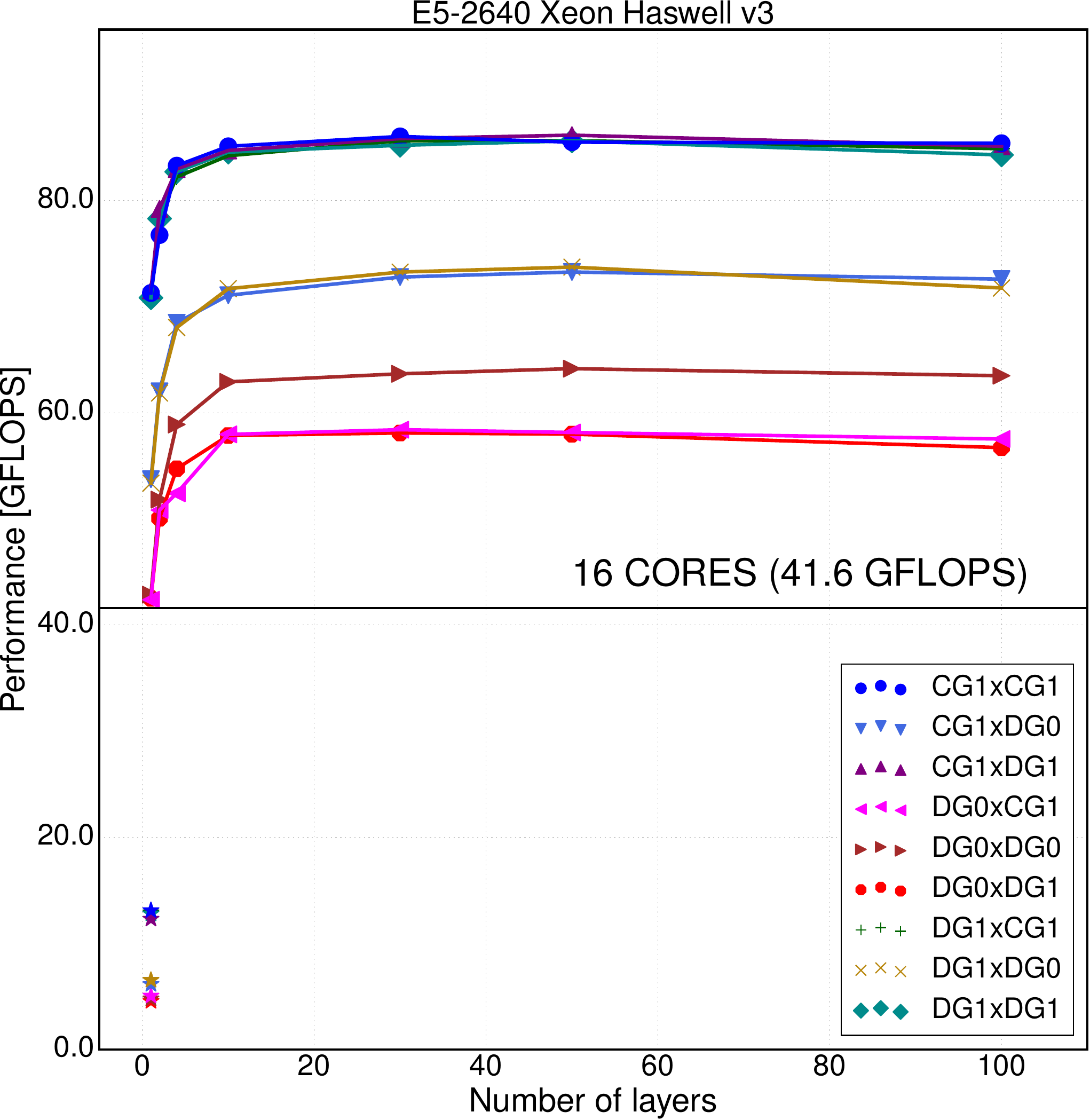}
  \caption{Performance of the $I$ integral computation on different
    data discretisations with varying number of layers on the Haswell
    architecture for a well-ordered base mesh. The star-shaped markers
    show the performance of the 1-layer badly-ordered mesh for
    comparison.  The horizontal line is the base FLOP throughput for
    $f_b = f_v = 1$ and the number of physical cores
    used.}
  \label{fig:haswell-good-ordering}
\end{figure}

\subsection{Discussion}
\label{ssec:discussion}

The performance of the extruded mesh iteration is constrained by the
properties of the mesh and the kernel computation.  The total number
of computations is based on the number of degrees of freedom per
cell. The range of discretisations used in this paper
(\autoref{fig:tpe-data-layout}) leads to four cases: one, two, three
or six degrees of freedom per cell. In compute bound situations,
discretisations with the same number of computations have the same
performance, see \autoref{fig:sandy-bridge-good-ordering} and
\autoref{fig:haswell-good-ordering}).

\subsubsection{Temporal locality}
\label{sssec:temporal-locality}
The numbering algorithm ensures good temporal locality between
vertically aligned cells. Any degrees of freedom which are shared
vertically are reused when the iteration algorithm visits the next
element. The reuse distance along the vertical is therefore minimal.

For $\mathrm{CG}$ discretisations, where degrees of freedom are shared
horizontally with other vertical columns, the overall performance
depends on the ordering of cells in the base mesh. Assuming a perfect
ordering of the base mesh, the numbering algorithm ensures a minimal
reuse distance while guaranteeing a minimum number of indirect
accesses and satisfying all the previously introduced spatial and
temporal locality requirements.

\autoref{fig:sandy-bridge-good-ordering} and
\autoref{fig:sandy-bridge-bad-ordering} demonstrate the combined
impact of horizontal mesh ordering and extrusion. In the extreme case
the flop rate increases up to 14 times between the badly ordered
single-layer case and the 100 layer well ordered case. This is
consistent with the widely held belief that unstructured mesh models
are an order of magnitude slower than structured mesh models.

The difference between well- and badly-ordered mesh performance
outlines the benefits responsible for the boost in
performance. Horizontal data reuse dominates performance for low
number of layers while spatial locality and vertical temporal locality
(ensured by the numbering and iteration algorithms) are responsible
for most of the performance gains as the number of layers increases.

We note, once again, that these results are for the lowest order
spaces which represent a worst case.  Higher-order methods both access
more contiguous data in each column and require many more FLOPs.  As a
result, we would expect to reach performance plateaus at lower numbers
of layers.

\section{Conclusions}
\label{sec:conclusions}
In this paper we have presented efficient, locality-aware algorithms
for numbering and iterating over extruded meshes. For a sufficient
number of layers, the cost of using an unstructured base mesh is
amortized. Achieved performance ranges from 70\% to 90\% of our best
estimate for the hardware's performance capabilities and current level
of kernel optimisation.  Benefits of spatial and temporal locality
vary with number of layers: as the number of layers is increased the
benefits of spatial locality increase while those of temporal locality
decrease.

This paper employed two simplifying constraints: that there are a
constant number of layers in each column, and that the number of
degrees of freedom associated with each entity type is a
constant. These assumptions are not fundamental to the numbering
algorithm presented here, or to its performance.  We intend to relax
those constraints as they become important for the use cases for which
Firedrake is employed.

The current code generation scheme can be extended to include
inter-kernel vectorization (an optimisation mentioned in
\citet{Meister:2015}) for the operations which cannot be vectorised at
intra-kernel level. The efficiency of such a generic scheme applicable
to different data discretisations is currently being explored.

In future work we intend to generalize some of the optimisations which
extrusion enables for both residual and Jacobian assembly:
inter-kernel optimisations, grouping of addition of contributions to
the global system and exploiting the vertical alignment at the level
of the sparse representation of the global system matrix. In addition
to the CPU results presented in this paper, we also plan to explore
the performance portability issues of extruded meshes on Graphical
Processing Units and Intel Xeon Phi accelerators.

\appendix
\paragraph{Acknowledgements}
This work was supported by an Engineering and Physical Sciences
Research Council prize studentship [Ref.\ 1252364], the Grantham
Institute and Climate-KIC, the Natural Environment Research Council
[grant numbers NE/K006789/1, NE/K008951/1, and NE/M013480/1] and the
Department of Computing, Imperial College London. The authors would
like to thank J. (Ram) Ramanujam at Louisiana State University for the
insightful discussions and feedback during the writing of this
paper. We are thankful to Francis Russell at Imperial College London
for the feedback on this paper.

\paragraph{Code availability}
The packages used to perform the experiments have been archived using
Zenodo: \cite{zenodo_firedrake}; \cite{zenodo_petsc};
\cite{zenodo_petsc4py}; \cite{zenodo_fiat}; \cite{zenodo_ufl};
\cite{zenodo_ffc}; \cite{zenodo_pyop2}; and
\cite{zenodo_coffee}. The source code repositories as well as the
archived versions are publicly available.

\paragraph{Data availability}
The scripts used to perform the experiments as well as the results are
archived using Zenodo: Sandy Bridge \citep{bercea_2016_61920} and
Haswell \citep{bercea_2016_61919}. The meshes used in the experiments
are available also \citep{bercea_2016_61819}. The archives are
publicly available.

\paragraph{Author contributions}
Gheorghe-Teodor Bercea designed the generalized extrusion algorithm,
performed the extension of the Firedrake and PyOP2 packages to support
extruded meshes, the performance evaluation and the preparation of the
graphs and tables. Andrew T. T. McRae extended components of the
Firedrake toolchain to support the finite element types used in the
experiments, and made minor contributions to the extruded mesh
iteration functionality. David A. Ham was the proponent of a
generalized extrusion algorithm. Lawrence Mitchell, Florian Rathgeber
and Fabio Luporini developed related features and framework
improvements in Firedrake, PyOP2 and COFFEE. Luigi Nardi is
responsible for the use of the floating point balance metric. David
A. Ham and Paul H. J. Kelly are the principal investigators for this
paper. Gheorghe-Teodor Bercea prepared the manuscript with
contributions from all the authors. All authors contributed with
feedback during the paper's write-up process.

\end{document}